\documentclass[12pt]{aa}
\usepackage{graphicx,natbib}
\usepackage{amsmath}
\usepackage{amssymb}
\usepackage{epstopdf}
\usepackage{xfrac}
\usepackage{txfonts}

\def\ftot{erg\,cm$^{-2}$\,s$^{-1}$}

\defcitealias{SchwopeEtAl2015-CSS-eph}{S+15}
\defcitealias{ScargleEtAl2013}{BB13}

\newcommand{\XMM}{\emph{XMM-Newton}}
\newcommand{\XSPEC}{{\sc XSpec}}
\newcommand{\SAS}{{\sc SAS}}
\newcommand{\CSS}{J071126}
\newcommand{\SEPH}{\citetalias{SchwopeEtAl2015-CSS-eph}}
\newcommand{\SEPHp}{\citepalias{SchwopeEtAl2015-CSS-eph}}
\newcommand{\pn}{EPIC-$pn$}

\newcommand\TopStrut{\rule{0pt}{2.6ex}}       
\newcommand\BottomStrut{\rule[-1.2ex]{0pt}{0pt}} 

\begin{document}

\title{\emph{XMM-Newton} and optical observations of the eclipsing polar
  CSS081231:071126+440405\thanks{Based on observations obtained with XMM-Newton, an ESA science mission 
with instruments and contributions directly funded by 
ESA Member States and NASA}} 
\date{\today}
\author{
H.~Worpel,
A.~D.~Schwope
}
\institute{Leibniz-Institut f\"ur Astrophysik Potsdam (AIP), An der Sternwarte 16, 14482 Potsdam, Germany}

\abstract{}
{We aim to study the temporal and spectral behaviour of the eclipsing polar 
CSS081231:071126+440405 from the infrared to the X-ray regimes.}
{We obtained phase-resolved \XMM\ X-ray observations on two occasions in
  2012 and 2013 in different states of accretion. In 2013 the \XMM\ X-ray and UV data were 
complemented by optical photometric and spectroscopic observations.}
{CSS081231 displays two-pole accretion in the high state. The magnetic fields
  of the two poles are 36 and 69 MG, indicating a non-dipolar
  field geometry. The X-ray spectrum of the main accreting pole with the lower
  field comprises a hot thermal component from the cooling accretion plasma, $kT_{plas}$ of a few
  tens of keV, and a much less luminous blackbody-like component from the accretion area with $kT_{\rm bb}
  \sim$ 50-100\,eV. The high-field pole, which was located opposite to the
  mass-donating star, accretes at a low rate and has a plasma temperature of
  about 4\,keV. On both occasions the X-ray eclipse midpoint precedes the
  optical eclipse midpoint by 3.2 seconds. The centre of the X-ray bright phase
  shows accretion-rate-dependent longitudinal motion of $\sim$20 degrees. }
{ CSS081231 is a bright polar that escaped detection in the RASS survey because it was in a low accretion state.
Even in the high state it lacks the prominent soft component previously thought to be ubiquitous in polars. Such an
excess may still be present in the unobserved extreme ultraviolet. All polars discovered in the \XMM\ era
lack the prominent soft component. The intrinsic spectral energy distribution of polars still awaits
characterisation by future X-ray surveys such as eROSITA. The trajectory taken by material to reach the
second pole is still uncertain.}

\titlerunning{ Observations of \CSS}
\maketitle

\section{Introduction}

Polars, also known as AM Herculis stars, are accreting binary systems in which material is transferred 
via Roche lobe overflow from a dwarf star onto a strongly magnetic white dwarf (WD). 
In these cataclysmic variable (CV) systems, no accretion disc forms because the WD's magnetic field is strong enough to cause the material lost by the secondary 
to travel down the magnetic field lines directly onto the magnetic poles. These systems are important 
for understanding accretion processes in the presence of a powerful magnetic field. 

An identifying characteristic of polars has historically been their prominent
soft X-ray emission. This property led to the discovery of numerous polars
with the EINSTEIN, EXOSAT, ROSAT, and EUVE satellites. However,
\XMM\ observations of X-ray selected polars (e.g. \citealt{RamsayCropper2004}) 
and the discovery of new polar systems reveal growing evidence that a soft X-ray
component is not present in the majority of polars.

CSS081231:071126+440405, henceforth \CSS, is a polar discovered on 2008
December 31 by the Catalina Sky Survey \citep{DrakeEtAl2009, TempletonEtAl2009}, when it suddenly
brightened by over three magnitudes. It had been observed previously, but
not recognised as a variable star. It is an eclipsing system with a period of
7031 seconds and an optical eclipse duration of 433.08 seconds
(\citealt{SchwopeEtAl2015-CSS-eph}; hereafter S+15).

Eclipses in polars benefit the observer because the viewing geometry is
constrained through knowledge of the inclination, and the periodic obscuration
of different sites in the system allows the contributions to the total flux
from the different emission sites to be distinguished. Additionally, if there are
any circumbinary planets, they can be detected by the R{\o}mer
delay of the eclipse (e.g. \citealt{SchwarzEtAl2009, QianEtAl2011, SchwopeThinius2014} for the case of HU Aquarii). 

\SEPH\ compiled a data set of optical observations of \CSS\                             
covering a time span of 5.3 years. They used the data to
obtain a precise linear ephemeris for the eclipse midpoints, finding no
evidence of any deviations from it, which is a result that sets a tentative upper 
limit on the mass of a circumbinary planetary companion of two Jupiter masses. 

\SEPH\ were unable to find a binary system accretion geometry that satisfied all of
their observational constraints. Though the high and intermediate accretion
states can be explained by a colatitude and azimuth of the magnetic axis of
$\beta = 18^\circ$  and $\psi=-3^\circ$, and a system inclination of
$i=79.3^\circ-83.7^\circ$, the location of the accretion spot in the low state
($\psi\approx 10^\circ$) is not consistent with these parameters. The authors
postulated a non-dipolar field structure as a possible solution. 

Observations of \CSS\ in ultraviolet and X-rays may help uncover the accretion
geometry. We present a study of two \XMM\ observations of the source, taken
one year apart, of \CSS\ in a high state (2013) and a state of reduced
accretion (2012). The observation in 2013 was accompanied by four nights of 
photometric observations with the STELLA robotic telescope \citep{StrassmeierEtAl2004}, 
and two nights of low-resolution spectroscopy at Calar Alto. 

In Sect.~\ref{sec:data_analysis} we give the details of our data reduction and
analysis methods. Sect.~\ref{sec:results} contains our results, and in
Sect.~\ref{sec:discussion} we discuss the possible accretion scenario at the two magnetic poles.

\begin{figure*}
 \includegraphics{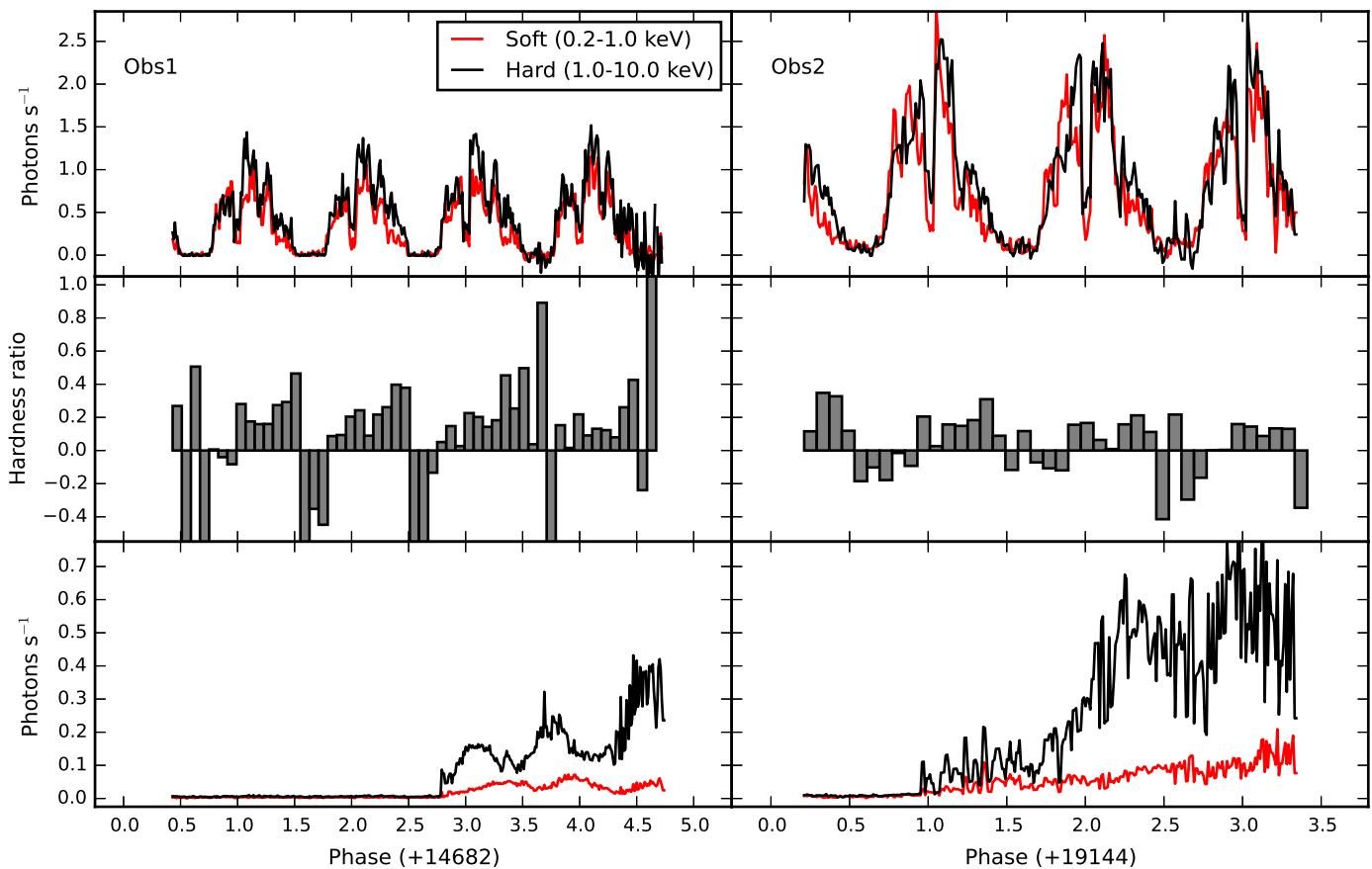}
 \caption{X-ray light curves of \CSS\ in both observations, and stacking  data for
   the EPIC-MOS1, MOS2, and $pn$ instruments. Top panels show background-subtracted
   source light curves in soft and hard X-rays. The middle panels show hardness ratios (Eq. \ref{eqn:hr}).
   Bottom panels show  exposure-corrected background light curves. The light 
   curves have been binned into 100 bins per cycle (70.3~s). The count rates for the 
   top two panels have been corrected for the enclosed count fraction of the source 
   extraction regions, but we have not corrected for data gaps in \pn. Two dips are visible in each of the first two cycles of Obs2 in
   soft X-rays.}
 \label{fig:raw_lightcurves}
\end{figure*}

\section{Observations and data reduction}
\label{sec:data_analysis}

\subsection{XMM-Newton X-ray and ultraviolet observations}
The \XMM\ satellite made two observations of \CSS. The first of these
(henceforth Obs1) was made on 2012 April 10 (observation ID 0675230101) and
lasted $\sim$32~ks. The second (Obs2) was made on 2013 April 8 (observation
ID 0693580101) and lasted $\sim$24~ks. The observations thus covered 4.59 and
3.42 cycles of the 1.95~hour binary. The \pn\ and -MOS cameras were operated in
full frame mode with the thin filter in both observations. \XMM\ observed
the star simultaneously with the Optical Monitor (OM) in fast mode with the
UVM2 and UVW1 filters for both observations. The UVM2 and UVW1 filters have
effective wavelengths of 2310\AA\ and 2910\AA, respectively \citep{KirschEtAl2004}. In Obs1
a total of 1 3197~s of UWV1 data and 1 5834.5~s of UVM2 data are available. In Obs2 we have
7 678~s and 14 169.5~s of data in those filters. We have not used the RGS
because of low signal-to-noise.

The raw data were reduced using the \emph{XMM-Newton} Science Analysis System
(\SAS) version 14.0.0. The \pn\ and MOS data were processed with the standard
tasks \emph{epchain} and \emph{emchain} to generate calibrated event
lists. The \emph{epreject} task was run for \pn\ data. All timing data in this
work were corrected to the solar system barycentre using the
\SAS\ \emph{barycen} task. The photon lists were filtered to only include
photons with energies between 0.15~keV and 12.0~keV for the MOS instruments
and between 0.15~keV and 15.0~keV for the \pn\ instrument. For the analysis of
hardness ratios and spectra we restricted the energies further (see subsequent sections). For each photon
arrival time we additionally assigned an orbital phase in the range $\phi \in [0,1)$ 
according to the ephemeris in \SEPH. Background-subtracted light curves were produced
with the \emph{epiclccorr} task to correct for vignetting, bad pixels, etc.
The OM data were reduced using the \emph{omfchain} task. 

Our source extraction regions were circles centred on the source, where the
precise location of the source was determined with the
\SAS\ \emph{edetectchain} task and the best extraction radius found with the
\emph{eregionanalyse} task. The source extraction radii were in the range 20 to
28 arcseconds, with enclosed energy fractions of over 90\% in each case. The
background extraction regions were boxes lying as near the source as was
practical, with care taken to ensure approximately the same CCD readout time.
We excluded a circular area of radius 42.5 arcseconds, centred on the source,
from the background-extraction region. 

\subsection{Optical photometry}
\CSS\ was observed with the robotic telescope STELLA during four nights in
2013 April. The overall settings, the data reduction and analysis, and the mean
light curve have already been presented in \SEPH. We compared the relative flux between 
our target and the comparison star \#139 of the American Association of Variable Star Observers (AAVSO) sequence.
The AAVSO gives a Henden sequence for this star with $B=14.696$ and $V=13.918$. 
STELLA observations were obtained through an SDSS g filter. The equation
given by \cite{BilirEtAl2005} was used to transform to the SDSS system, $g_{\#139} = 14.3$.
 
STELLA observations were performed on the nights of April 1, 7, 10, and 12. Full 
phase coverage was achieved on each occasion. The optical
observations that happened closest to the \XMM\ observations began on
April 7 and covered orbital phases 19135.20 -- 19137.91, so only a few cycles earlier than
the X-ray observations beginning in cycle 19144. 

\subsection{Optical low-resolution spectroscopy}
Low-resolution spectroscopy was performed during the nights of 2013 March 20 and March
21 with the 2.2m telescope of the Calar Alto
observatory. It was equipped with CAFOS, a grism spectrograph and imager. The 
B400 grism provided a wavelength coverage from 3360 -- 9650\,\AA\ at a FWHM 
resolution of 9.7\,\AA\ as measured from calibration lamp spectra. 
The integration time per spectrum was 3\,m. The time resolution thus achieved
was 243\,s (0.035 phase units). On both nights 19 spectra were taken, but
clouds at the end of the second observation degraded the quality of the
spectra considerably.
The spectroscopic observations covered the binary cycles 18913.96 --
18914.62 and 18925.59 -- 18926.18. Although these observations were a few weeks earlier 
than the STELLA and \XMM\ observations, the bottom panel of Fig. \ref{f:mwlcs} shows that \CSS\ was similarly
bright in the optical on both occasions. We are therefore confident that we are comparing similar accretion 
states.

The spectrograph was rotated so that comparison star \#142 from the AAVSO
chart could be observed simultaneously. This star at RA2000 = 7:11:35.5,
DE2000 = 44:04:03.3 has $B=15.167, V=14.158, Rc=13.621, Ic=13.109$. 

Arc lamp spectra (Hg+He+Rb) for wavelength calibration were obtained before
and after the sequence of the target star. No spectrophotometric standard
star was observed during those two nights through the same grisms. To
achieve a rough spectrophotometric calibration, standard star observations
were sought in the data archive at Calar Alto. We found two standard 
star observations with the same instrumental setup. 
The two spectral response curves, based on observations of the standard  stars
BD+28\_4211 observed in May 2010 and HZ44 observed in March 2008,
 differed by a factor of almost 2 in absolute flux but showed
a similar shape. The response curves were thus used to put the target spectra on a
relative scale. 

To achieve a proper photometric calibration and to put the spectra on an absolute scale 
the data of the comparison star \#142 were used. Firstly, an achromatic correction 
with the mean brightness of the comparison star in the wavelength interval 5000 -- 7000\,\AA\
was applied to the spectra of both the target star and the comparison star. This procedure accounted 
for seeing and transparency variations. Secondly the average spectrum of the comparison star
was folded through the BVRcIc filter curves and correction factors determined 
at the corresponding central wavelengths of the filters. The factors were then fitted with a polynomial 
of degree 2 and the correction function was applied to all spectra of the target. The correction 
function is not fully determined at the blue and red ends of the spectrum owing to 
the lack of sampling points. Given the uncertainties of the various reduction 
steps, we estimate a photometric accuracy of our final spectra 
of about 15\% in the centre and of 30\% at the end of the spectral range covered
with the B400 grism.

\section{Analysis and results}
\label{sec:results}

\subsection{X-ray light curves}
\label{sec:xray_lightcurves}
\begin{figure}
 \includegraphics{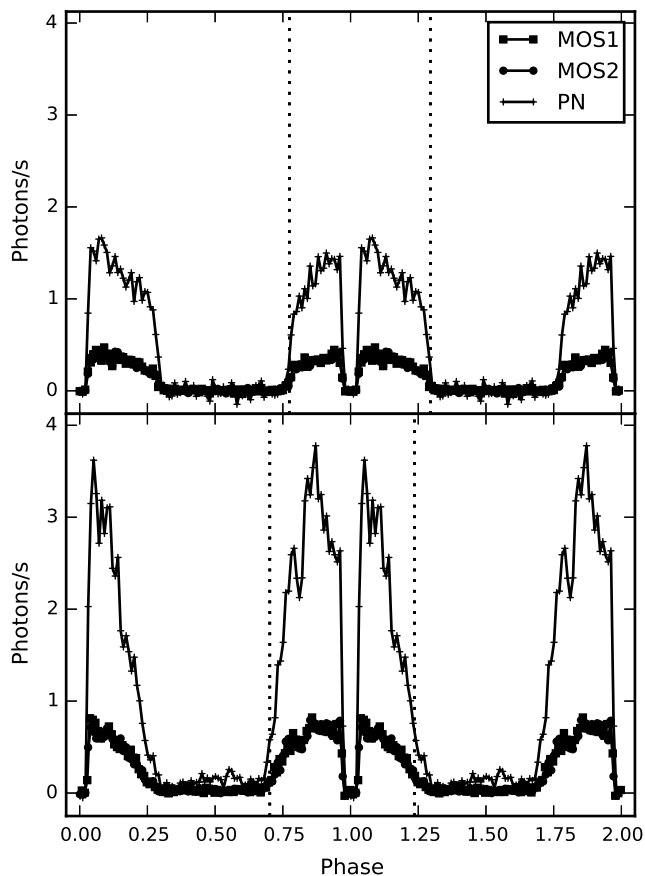}
 \caption{Phase-folded X-ray light curves of \CSS\ in both observations, for
   the EPIC-MOS1, MOS2, and $pn$ instruments. The light curves have been
   binned into 100 bins per cycle, and two phases are plotted for clarity.
   The energy ranges are 0.2-12~keV for the EPIC-MOS instruments and 0.2-15~keV for \pn, and
   count rates have been corrected for the enclosed count fraction of the
   source extraction regions. Dotted vertical lines indicate the beginning and end
   of the bright phase, clearly showing the longitudinal shift.} 
 \label{fig:phasefolded_lightcurves} 
\end{figure}

Figure \ref{fig:raw_lightcurves} shows the X-ray light curves of the two
observations. As is typical of eclipsing polars, the light curves are
dominated by the emission from a self-eclipsing accreting pole, interrupted by
a short period of very low flux when the WD and main accreting pole are
eclipsed by the secondary star. The X-ray flux is larger in
the second observation, indicating a higher accretion rate. The
bottom two panels of Fig.~\ref{fig:raw_lightcurves} show the background
flux. Soft proton flares begin to affect both observations about half way
through, and Obs2 is more severely affected.

Phase-folded light curves are shown in
Fig.~\ref{fig:phasefolded_lightcurves}. The shape of the light curves suggests that the bright phases have shifted in longitude between Obs1 and Obs2. To 
verify this observation, we found the centres of the bright phases as follows. We binned the data into 100 bins per phase. 
We defined the peak count rate of the bright phase as the mean of the brightest three
points in each cycle, for each instrument, and took the rise and fall of the 
bright phase to be the times at which the count rates reached or dropped below 
20\% of the peak count rate. Taking unweighted averages for each cycle, we found the
bright phases were centred at $\phi_c=0.035\pm 0.003$ and $\phi_c=-0.031\pm 0.006$ for Obs1 and Obs2.

This result shows that the midpoint of the bright phase is indeed earlier in 
Obs2 than in Obs1. These bright phase midpoints correspond to a trailing
longitude of $\psi\approx-11\pm 2^\circ$ in Obs1 and a leading spot longitude of
$\psi\approx12\pm 1^\circ$ in Obs2. Some care should be taken in interpreting these numbers
because the bright phase is asymmetrically shaped, and therefore the exact midpoint
can depend on what count rate is taken to mark its rise and fall. Nonetheless, the motion of 
the bright spot between observations is qualitatively similar to the accretion-rate dependent
motion found for this source in optical light curves in \SEPH.

There is a slight indication in Fig. \ref{fig:phasefolded_lightcurves} of an increase 
in X-rays around phase $\phi\approx 0.5$ in Obs2, suggesting a second accreting pole. We discuss this
secondary hump in Sect. \ref{sec:secondhump}.

\begin{figure}
 \includegraphics{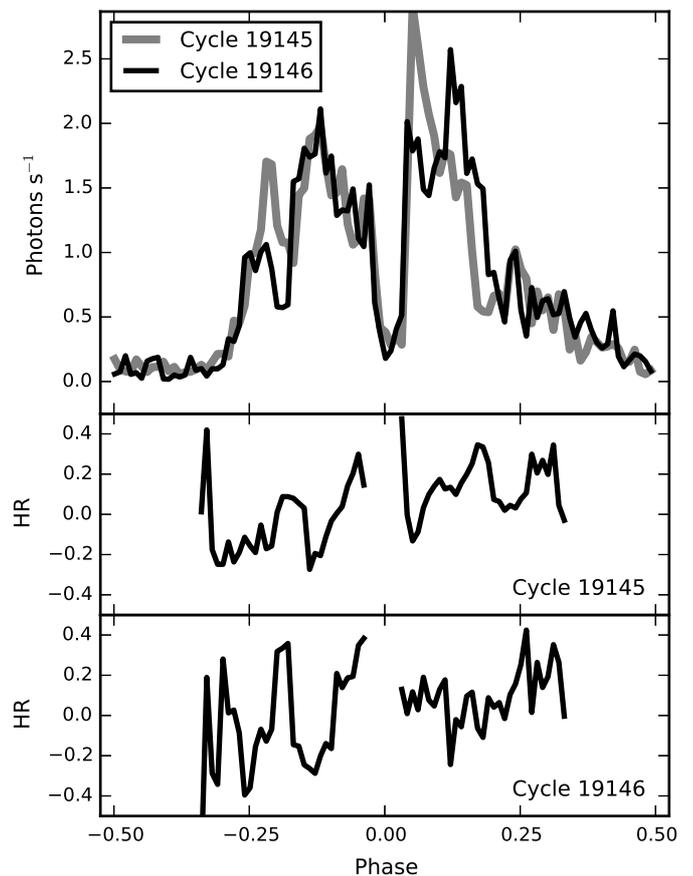}
 \caption{Cycles 19145 and 19146 (Obs2) in soft (0.2-1.0 keV) X-rays (top panel). Bright phase dips occurring at phase $0.81-0.83$
 and pre-eclipse dips beginning at about phase 0.9 and leading into ingress are visible. The middle and bottom panels show 
 hardness ratios (Eq. \ref{eqn:hr}) for these
 two cycles. The two dips are much harder than the rest of the bright phase. For clarity the eclipses and
 faint phases are not plotted in the hardness ratio panels. The third cycle of Obs2 is omitted because it is severely affected by data gaps.}
 \label{fig:obs2_soft}
\end{figure}

We have also analysed hardness ratios in X-rays for both observations:
\begin{equation}
 \text{HR}=(H-S)/(H+S),
 \label{eqn:hr}
\end{equation}
where $H$ and $S$ are the background-corrected count rates in the 1.0-10.0~keV
energy band and the 0.2-1.0~keV energy band. The choice of non-standard 
bands for the hardness ratio was motivated by the presence of
two radiation components in the soft X-ray regime below 1\,keV and a 
hard thermal spectrum above this energy (see Sect. \ref{sec:xray_spectroscopy}). 
The results are shown in the middle panels of Fig.~\ref{fig:raw_lightcurves}. 
For Obs1 the hardness ratio is around $\text{HR}\approx0.2$ for the bright phase 
and much softer in the faint phase, and there is slight evidence that
the bright phase has a rising hardness profile. In Obs2, the bright phase has a hardness
ratio of 0.1, and the faint phase is also softer.

The soft X-rays from the first two
cycles of Obs2 are shown in Fig. \ref{fig:obs2_soft}. Aside from the eclipse, two features are evident:
a dip at phases 0.81 and 0.83 (the \emph{bright phase dip}) and another just before eclipse ingress (the \emph{pre-eclipse dip}).
Both bright phase dips have been observed previously in, for example, HU Aqr \citep{SchwopeEtAl2001}, but in this source they are narrower and
occur later in phase. The bright phase dip was previously found in optical for this source by \cite{KatyshevaShugarov2012}. 
Both features have higher hardness ratios than the rest of the bright phase, suggesting that they arise from absorption, 
which primarily affects the low energy end of the X-ray spectrum. It also seems that the bright phases have a
rising hardness ratio profile in both observations.

\begin{figure}
 \includegraphics{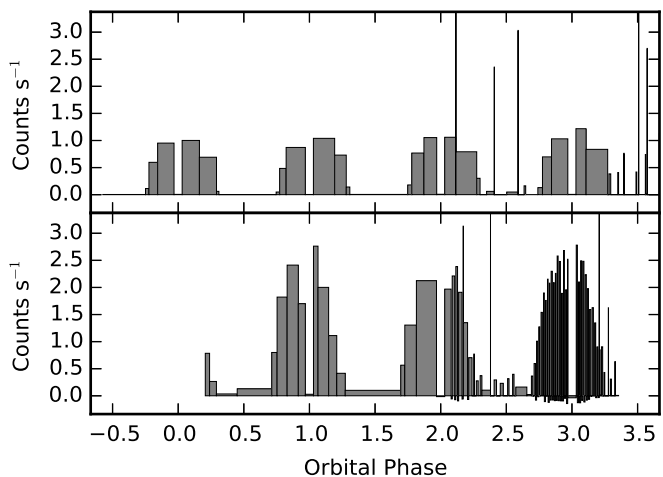}
 \caption{Bayesian Block light curves of Obs1 and Obs2. The eclipses are clearly defined, 
 except for the third eclipse of Obs2, which is affected by data gaps. The gaps
 are plainly visible, beginning around phase 2.1. A few noisy
 blocks of implausible count rate are also visible (see \citealt{WorpelSchwope2015}) but do not
 affect the eclipse timings.}
 \label{fig:lightcurves}
\end{figure}

We applied the Bayesian Block method \citep{Scargle1998,ScargleEtAl2013} to determine the starts and
ends of the eclipses; see also \cite{SchwopeEtAl2002}
for an earlier application of this algorithm to CV eclipse timing. Because we
need to perform background subtraction, with the data degraded by soft proton
flares that affect the second half of both observations, we have used the
weighted-photon adaptation to the original algorithm described in
\cite{WorpelSchwope2015}. We produced Bayesian Block representations for all
instruments but only \pn\ data were used for the timings because the EPIC-MOS
data is accurate to only 2.6~s in full-frame mode. The \pn\ instrument has
a timing resolution of 0.0734~s \citep{KirschEtAl2004b}. The Bayesian Block light
curves are shown in Fig.~\ref{fig:lightcurves}. These do not contain corrections for vignetting, bad pixels, etc.
which for normal light curves would be performed by the \emph{epiclccorr} task. This is why the count rate appears lower.
The eclipse parameters are
given in Table \ref{tab:eclipse_timings}. Uncertainties on a Bayesian Block
change point was determined by generating many simulated sequences of photon
arrival times with the same count rates as the blocks the change point
separates, for source region and background region photons. The third cycle of Obs2 is
affected by periodic data gaps in \pn\ 
data because the total count rate exceeds telemetry limits (see \emph{XMM-Newton Users Handbook,
} Sect. 4.3.1). The data gaps begin in the fall of the second bright phase and
prevent accurate eclipse timing for the third eclipse, so for the third eclipse we used 
EPIC-MOS2 data, which are unaffected but have poorer timing resolution.

\subsection{Multifrequency light curves}
\label{sec:OM}

\begin{figure}
\includegraphics{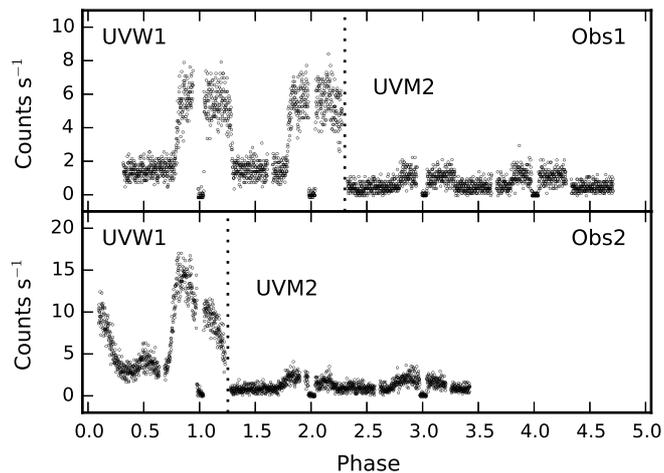}
\caption{Optical monitor light curves for the two observations as a function of orbital phase. The binning is 10~s.
For both Obs1 and Obs2 the filters were changed from UVW1 to UVM2 partway through the 
observation, as indicated by the dotted vertical lines.}
\label{fig:OM_lightcurves}
\end{figure}

\begin{figure}
\includegraphics{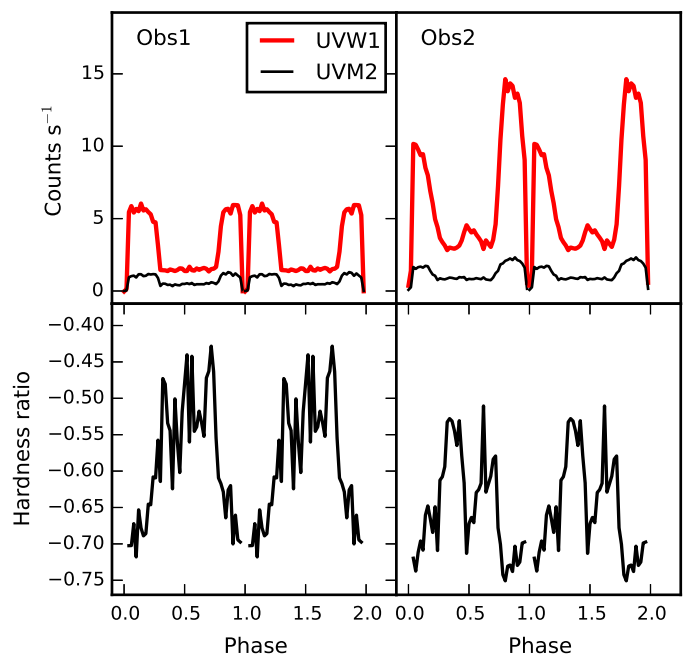}
\caption{Phase-folded optical monitor light curves (top panels) and hardness ratios (bottom panels).
The secondary hump in Obs2 is visible in UVW1, but is not apparent in UVM2. 
The bright phase is much softer than the faint phase, as is the secondary hump. Obs2 is
generally slightly softer than Obs1.}
\label{fig:OM_pf_lightcurves}
\end{figure}

The OM light curves (see Figs. \ref{fig:OM_lightcurves} and
\ref{fig:OM_pf_lightcurves}) are qualitatively similar to the X-ray light
curves. In both observations the accretion spot is plainly visible as a bright
phase, which is interrupted by a deep eclipse. In Obs1, the bright phase is symmetrical
and flat-topped. There are no visible changes in its shape over two orbital
cycles in either filter. In Obs2 the bright phase has a sharp rise and slow
decline in both filters, though the rise is visibly slower in the UVM2
filter. Additionally, in Obs2 there is a small secondary hump at phase
$\phi\approx 0.55$, but only in the softer filter. There is no evidence of this
in the UVM2 data, but this may simply be due to the lower sensitivity of that filter.
The secondary hump also has a fast rise and slow decline, extending from about
$\phi=0.45$ to $\phi=0.65$, but determining its end is difficult because of a
data interruption during the UVW1 observation. 

There is no evidence of any pre-eclipse or bright phase dips in either observation. This is
surprising, since the features were present in soft X-rays in Obs2. Even a low $N_H$ column
density of $5\times 10^{20}$~cm$^{-2}$ (see Sect. \ref{sec:xray_lightcurves}) would cause a factor 2
extinction at 2500\AA, assuming the absorption properties of local absorbing material is similar to
interstellar material. However, in this part of the spectrum absorption
is dominated by free-free absorption and the optical depth is proportional to the square of the
column density \citep{WatsonEtAl1989, WatsonEtAl1990}. Thus even a slightly lower $N_H$ would result in a much more subdued
dip feature.

As with the X-ray hardness ratios, there is clear evidence of a phase
dependence on the UV hardness ratio. The bright phase is softer than the faint
phase in both observations, opposite to the behaviour in X-rays. The secondary hump in Obs2 is
also softer in UV than the faint phase. Only the qualitative behaviour of the hardness
ratio is relevant, not the values themselves, because of the different transmission properties
of the UVW1 and UVM2 filters.

The midpoint of the bright phase moves from a trailing
longitude in Obs1 to a leading longitude in Obs2, similar to the behaviour of
the bright phase in X-rays. Taking the rise and fall of the bright phase to be
20\% brighter than the mean faint phase brightness, and stacking the phase-folded light curves 
and excluding the secondary hump in Obs2, we found a trailing spot longitude of $\psi\approx -10^\circ$ in
Obs1 and a leading spot longitude of $\psi\approx 10^\circ$ in Obs2.

\CSS\ was gradually becoming brighter by some 60\% in the first
half of 2013 April as revealed through our STELLA/WiFSIP monitoring.
Mean g-band magnitudes in the bright ($\phi = 0.89 - 0.93$) and faint
($\phi = 0.31-0.35$) phases were found to be 15.12, 15.10, 14.94, 14.64
and 17.15, 17.14, 16.92, 16.66 on April 1, 7, 10, and 12, respectively.
On all occasions there was a steeper increase towards the orbital
maximum and a slower decline. There was a shoulder on the decreasing branch 
of the light curve at orbital phase 0.27. On April 12, when the source
appeared brightest, further shoulders occurred at phases 0.75 and 0.21, that
might indicate that the source develops further emission regions.

\begin{figure}
\resizebox{\hsize}{!}{
\includegraphics[clip=]{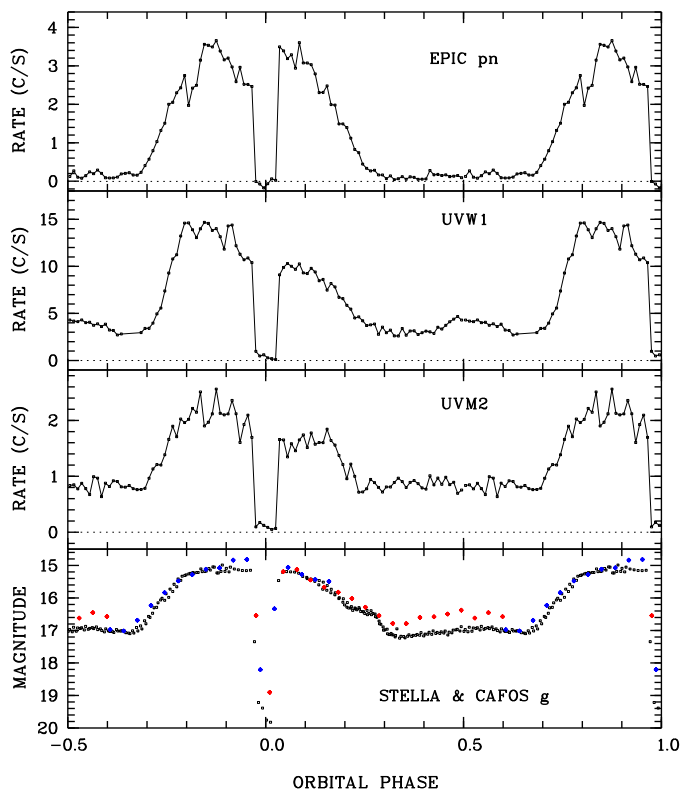}}
\caption{Phase-folded X-ray, ultraviolet and optical data obtained 2013 March \&
  April. The bin size is 0.01 phase units (70.3~s) in the three upper panels. The
  optical data shown in the lowest panel shows STELLA data (60~s bin size) obtained on the
  night of April 7/8 and light curves obtained from the low-resolution
  optical spectroscopy with CAFOS (red: March 20, blue March 21; 243~s bin size).}
\label{f:mwlcs}
\end{figure}

All multi-wavelength light curves obtained in 2013 April, which are averaged into 100
phase bins, are displayed in Fig.~\ref{f:mwlcs}. All have the asymmetric shape
in common, reaching maximum brightness during the first half of the bright
hump. The X-ray light curve appears to be the most symmetric of the four light curves shown. The steep increase and 
the slower decline of the low-energy light curves is likely due to the optical thickness
effects of the cyclotron component. The lowest panel also includes
g-band data from the CAFOS spectroscopy in March 2013.

\subsection{The main orbital hump}
\subsubsection{X-ray spectroscopy}
\label{sec:xray_spectroscopy}
During both observation X-ray spectra  were extracted using the bright phase
intervals, which were identified by visual inspection of the Bayesian Block light curves, and 
excluding the eclipses. We analysed the X-ray spectra using version
12.8.2p of the \XSPEC\ software package \citep{Arnaud1996, DormanArnaud2001}. During spectral
fitting the \pn\ and EPIC-MOS data were analysed simultaneously in an energy
range of 0.2$-$10.0~keV for \pn\ and 0.2$-$8.0keV for EPIC-MOS. To accommodate
bins with low photon counts, we used Churazov weighting \citep{ChurazovEtAl1996}.

\begin{figure}
 \includegraphics{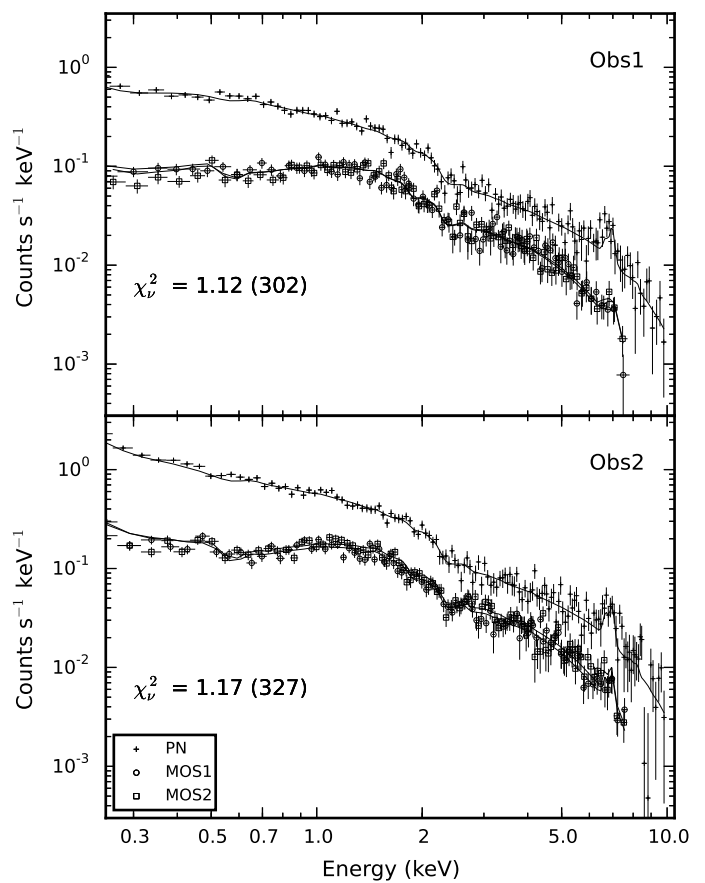}
 \caption{Simultaneous fits to \pn\ and EPIC-MOS spectra of the bright 
 phase of both observations. The data are shown as points with error bars, 
 and the fit to the data is represented by solid black lines. The spectra 
 are well fitted by the {\sc Mekal}+\emph{bbodyrad} model.}
 \label{fig:bright_phase_spectra}
\end{figure}

We initially fit the data for both observations using a single {\sc Mekal}
plasma model \citep{MeweEtAl1985,LiedahlEtAl1995}, with the metal plasma
abundances frozen at solar values and redshift at zero. These fits did not give good $\chi^2_\nu$
values. Visual inspection of the fits showed that the major
cause of the poor $\chi^2_\nu$ values was an excess of photons at low
energies. To correct this, we fitted the data again with
the sum of a {\sc Mekal} component and a soft blackbody. These fits gave adequate $\chi^2_\nu$. The spectral parameters are
given in Table \ref{tab:spectral_fits}. For a distance of 390~pc, the soft blackbody in Obs2 would represent a sphere of
radius 510~km. This value is plausible, since it is much smaller than the radius of the WD itself.

The uncertainties on the plasma temperature were found using \XSPEC's \emph{steppar} command to achieve a confidence interval of 99\%.
Uncertainties on fluxes for all models were calculated using 
the \texttt{cflux} convolution component. The uncertainties in the fluxes are often 
proportionally smaller than those on the temperatures and normalisations because these two parameters vary independently.
The spectral fits are shown in Fig.~\ref{fig:bright_phase_spectra}. The necessity of adding a second
component to the low-energy end suggests that $N_\text{H}$ absorption is
likely to be small. We confirmed this result by adding a \emph{tbabs} multiplicative
component to the {\sc Mekal}+\emph{bbodyrad} model and repeating the fits. 
We obtained an $N_H$ value consistent with zero and no improvement in $\chi^2_\nu$.

We determined upper limits on $N_H$ by increasing it until no statistically acceptable
fit could be obtained, to 99\% confidence. This upper limit was $0.135\times 10^{22}$~cm$^{-2}$ for Obs1 and
$0.081\times 10^{22}$~cm$^{-2}$ for Obs2. We found maximum unabsorbed fluxes for the {\sc Mekal} and
blackbody components by freezing $N_H$ at its upper limits, and the other spectral component parameters
to their previously determined values, and fitting again. The upper limits to the fluxes for the {\sc Mekal}
component were $4.2\times10^{-12}$~erg~s$^{-1}$~cm$^{-2}$ in Obs1 and $6.8 \times 10^{-12}$~erg~s$^{-1}$~cm$^{-2}$
in Obs2. For the blackbody component the maximum fluxes were $3.6\times 10^{-12}$~erg~s$^{-1}$~cm$^{-2}$ in
Obs1 and $6.1\times 10^{-12}$~erg~s$^{-1}$~cm$^{-2}$ in Obs2.

To estimate the amount of absorption in the two dip features we returned to the {\sc Mekal}+\emph{bbodyrad} model in Table \ref{tab:spectral_fits}
and added a \texttt{tbabs} absorption coefficient to determine the $N_H$ needed to produce a given reduction in 
flux between 0.2 and~1.0 keV. We required $1.4\times 10^{20}$, $4.3\times 10^{20}$, and $1.4\times 10^{21}$~atoms~cm$^{-2}$
to achieve reductions in flux of 25\%, 50\%, and 75\%. We have selected a range of flux reduction 
factors because the actual depths of the dips is difficult to estimate. Similar column depths have 
previously been reported for pre-eclipse dips in HU~Aqr \citep{SchwarzEtAl2009}.

Inspection of the spectral fits in Fig. \ref{fig:bright_phase_spectra} suggests the possible presence of a
fluorescent iron line at 6.4keV, based on four or five bins of positive residuals. We estimated the equivalent
width of the possible iron line as follows. We took the {\sc Mekal} + \emph{bbodyrad} model with the parameters of
Table \ref{tab:spectral_fits} frozen, and restricted the energy range to 5.0-8.0~keV to isolate the emission
line complex. Even in this energy range, the model is statistically acceptable. We added a Gaussian emission
line with energy frozen at 6.4~keV, but with width and normalisation allowed to vary, and fitted the spectrum again.
The fits suggest an equivalent width, using XSPEC's \emph{eqwidth} command, of 100-150~eV in both observations.
We then increased the normalisation of the Gaussian until the fit was no longer statistically acceptable (to 99\% confidence),
giving an upper limit of about 500~eV in both observations. These numbers are consistent with the results of \cite{EzukaIshida1999},
who found poorly constrained equivalent widths of similar magnitude for several polars using \emph{ASCA} data. We emphasize, however, that the evidence of the line being there at all is not strong.

To quantify any possible effect of the proton flares on the spectral fits, we extracted bright phase spectra for all instruments in each orbital
cycle separately and compared them to the corresponding model in Table \ref{tab:spectral_fits}. The four cycles of Obs1 gave
$\chi^2_\nu$ of 1.11, 1.08, 0.96, and 1.06. In Obs2 we obtained $\chi^2_\nu$ of 1.70, 1.94, and 1.16, but the poor results for the first two
could be remedied by allowing the temperature of the blackbody component to vary by $\sim10$~eV. Since the proton flare is strongest at
high energies (Fig. \ref{fig:raw_lightcurves}), it is likely that this result represents a definite variation in the soft component of the
spectrum. There is therefore no evidence that background subtraction of the soft proton flare is inadequate even for periods of strong flaring. We have therefore also 
not extracted good time intervals.

\begin{table}
 \caption{ Spectral fit parameters and their uncertainties, fit statistics, and model 
 bolometric fluxes for the bright phase in both observations. Question marks in the
 upper or lower uncertainties indicate that no constraint could be determined. Normalisations
 for the {\sc Mekal} model are given in units of $10^{-14} \int n_e n_HdV / 4\pi[D_A(1+z)]^2$, where
 $D_A$ is the angular diameter of the source, and $n_H,n_a$ are hydrogen and electron densities. Normalisations for
 the blackbody model are in units of $(R/D_{10})^2$ where $R$ is the radius of the object in kilometers and $D_{10}$ is its
 distance in units of 10~kpc. All fluxes are calculated in the energy range $10^{-6}-100.0$~keV. This energy range is wide enough that small
 changes in the upper and lower limits do not change the results.
 Uncertainties are given at the 99\% confidence level.}
 \label{tab:spectral_fits}
 \begin{tabular}{lll}
  \hline
 \multicolumn{3}{c}{ {\sc Mekal+Bbodyrad}, 0.2-10.0~keV } \\
 \hline
 & Obs1 & Obs2\\
 \hline
kT$_\text{mekal}$ (keV) & $17.0^{+6.6}_{-4.0}$&$14.1^{+3.7}_{-2.4}$\TopStrut\\
norm$_\text{mekal}$ ($10^{-3}$) & $1.49^{+0.08}_{-0.06}$&$2.49^{+0.11}_{-0.08}$\TopStrut\\
kT$_\text{bb}$ (keV) & $0.09^{+0.03}_{-0.03}$&$0.05^{+0.01}_{-0.01}$\TopStrut\\
norm$_\text{bb}$ & $175^{+1900}_{-125}$&$13000^{+18000}_{-7000}$\TopStrut\\
$\chi^2_\nu$ (dof) & 1.12 (302)&1.17 (327)\TopStrut \\

Flux ($10^{-12}$ erg s$^{-1}$ cm$^{-2}$) & $3.86^{+0.08}_{-0.08}$&$8.10^{+0.15}_{-0.15}$\TopStrut\\
Flux$_\text{mekal}$ ($10^{-12}$ erg s$^{-1}$ cm$^{-2}$) & $3.73^{+0.08}_{-0.08}$&$6.23^{+0.13}_{-0.13}$\TopStrut\\
Flux$_\text{bb}$ ($10^{-12}$ erg s$^{-1}$ cm$^{-2}$) & $0.13^{+0.03}_{-0.03}$&$1.27^{+0.12}_{-0.12}$\TopStrut\\

\end{tabular}
\end{table}

\begin{figure}
\resizebox{\hsize}{!}{
 \includegraphics[angle=270,clip=]{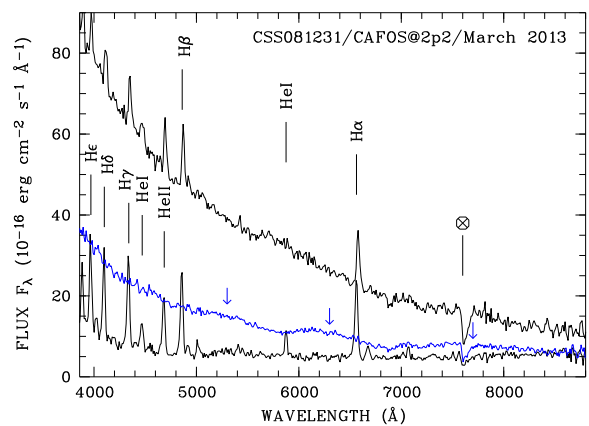}
}
 \caption{ CAFOS spectra obtained during bright ($\phi = 0.92 - 0.95$) and faint
phases ($\phi = 0.31 - 0.37$) a few weeks prior to Obs2.
The spectrum displayed in blue was obtained at phase 0.18 with
atomic emission lines subtracted. It shows modulations that were
identified with cyclotron harmonics indicated by arrows. Hydrogen Balmer, HeI, and HeII
 lines are prominent. The feature indicated with a cross is an atmospheric line; the spectra have not
 been corrected for them.
}
 \label{f:cafos}
\end{figure}

\begin{figure}
\resizebox{\hsize}{!}{
 \includegraphics[clip=]{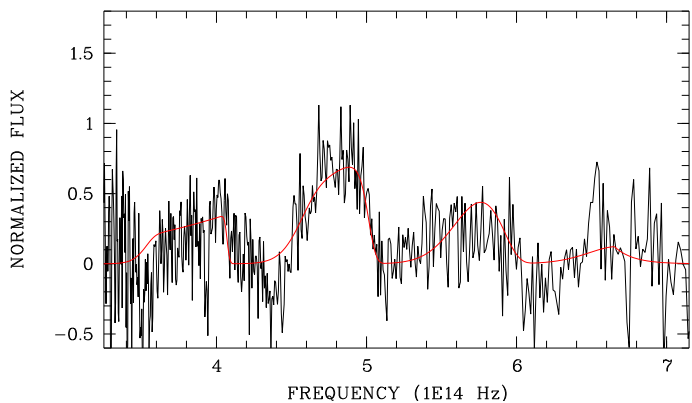}}
 \caption{Continuum-subtracted cyclotron humps observed at phase 0.18 with a cyclotron model for 36 MG superposed
}
 \label{f:cycmod1}
\end{figure}

\subsubsection{Optical spectroscopy}
Optical spectra representing the minimum phase and the maximum phase
are displayed in Fig.~\ref{f:cafos}. They show the typical emission line spectrum of polars in their 
high states with H-Balmer, HeI and HeII emission. The minimum spectrum shows a flat continuum,  
while the continuum in the bright phase is very blue with a maximum intensity outside the spectral range
covered here. The continuum does not show a completely smooth variation, and in particular the 
spectrum taken during the rise to and fall from the bright phase shows undulations that are reminiscent of cyclotron harmonics.

The harmonics offer the opportunity to measure the field strength in the main accretion region. To extract 
the cyclotron harmonic lines from the main pole, we first subtracted the main emission lines from each 
spectrum by fitting Gaussians interactively. We then subtracted the spectrum obtained at minimum phase
from the spectrum at fall phase and fitted a low-order polynomial to the difference, representing the 
continuum. After subtracting this continuum one is left with a spectrum that is modulated  by cyclotron 
harmonic lines and noise. The cyclotron harmonics are centred on 5200, 6300, and 7700\,\AA\ and can be 
reproduced with an isothermal cyclotron model assuming $kT = 5$\,keV and $B_1 = 36$\,MG (Fig.~\ref{f:cycmod1}) using the 
models in a series of papers \citep[see e.g.][and references therein]{CampbellEtAl2008, SchwopeEtAl2003}.

Cyclotron harmonic humps could be recognised only during the rise and fall of the bright phase but not 
in its central part. Interestingly, the cyclotron humps seem to be shifted as a function of phase. 
They are most blueshifted at the beginning and end of the bright phase and become redshifted during 
the bright phase, when the line of sight comes closer to the magnetic axis in the accretion spot.
This is opposite to what is observed in other polars where the highest redshift was observed 
when the accretion spot was approaching the limb of the WD \citep[see e.g.][for the cases of HU Aqr 
and RX\,J0453.4$-$4213]{SchwopeEtAl2003, BurwitzEtAl1996}, and it is contrary to what is expected for  
plasma emission from a point-like, isothermal accretion region. As a result, this observation could 
be evidence of an extended cyclotron-emitting region with successively self-eclipsing parts 
belonging to regions of slightly different field strengths. A mixture of different field strengths would
manifest as a blurring of the cyclotron harmonics (e.g. \citealt{SchwopeEtAl1990}), but our
spectroscopic data do not have high enough quality to investigate this effect.

The flux in the cyclotron component is difficult to determine from optical observations alone
since it peaks shortward of the spectral range covered by our low-resolution spectroscopy.
The similarity of the optical and ultraviolet light curves suggests they are formed by the same
emission process, predominantly cyclotron radiation. Photospheric radiation from the heated 
region around the accretion spot might also be present, but seems to be a minor contributor in this case, 
since photospheric radiation typically gives symmetric light curves \citep[e.g.][]{GaensickeEtAl1998}. The bright-phase peak flux in the optical (at
3500\,\AA) was $8.5 \times 10^{-15}${\,erg cm$^{-2}$ s$^{-1}$ \AA$^{-1}$} (corrected for
the faint-phase contribution) whereas it was $(5.2\pm0.5) \times 10^{-15}${\,erg cm$^{-2}$ s$^{-1}$ \AA$^{-1}$} and
$(3.1\pm0.5)\times 10^{-15}${\,erg cm$^{-2}$ s$^{-1}$ \AA$^{-1}$} in the UV at 2910\,\AA\ and 2310\,\AA, respectively.
The optical-UV SED therefore shows a peak at about 3300\,\AA\ that isteeply decreases
towards the UV and the optical. The shape is not
compatible with a hot photospheric spectrum from the WD but is
consistent with one dominated by cyclotron radiation. The bolometric
flux in this component is $F_{\rm cyc} \simeq  2.2\times10^{-11}$\,\ftot, so roughly 
three times more luminous than the X-ray plasma component.

\subsection{Optical and X-ray spectroscopy of the secondary orbital hump}
\label{sec:secondhump}
\begin{figure}
\includegraphics[angle=270, width=100mm]{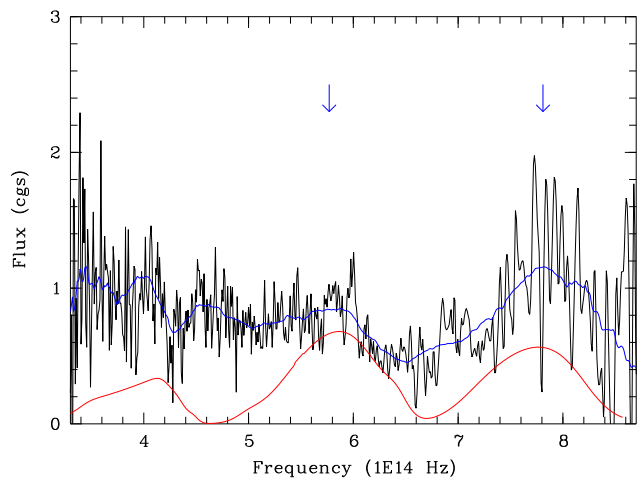}
 \caption{ Mean faint phase spectrum ($\phi = 0.38 -0.60$) after subtracting
atomic emission lines and the mean spectrum of the phase interval
$0.31-0.37$. The blue line is a boxcar-smoothed version of the same
data. The arrows indicate tentatively identified cyclotron harmonics.
The red line is a 3 keV cyclotron model from a 69 MG field.}
\label{f:cyc2pole}
\end{figure}

Optical light curves have occasionally shown a secondary hump centred on phase 0.50-0.60, as described in \SEPH. This 
feature is also present in the data obtained with the OM (see Sect. \ref{sec:OM}) between phases 
$0.45$ and $0.65$ in the UVW1 filter. The secondary hump is also visible in the g-band STELLA light curves, as 
well as in the CAFOS photometry (see Fig.~\ref{f:mwlcs}). The feature remained unexplained
by \SEPH. The new data allow us to study the origin of the secondary hump and test whether it could arise from accretion onto a second pole. 
We isolated an optical spectrum as follows. We obtained a spectrum of the orbital minimum, which occurs in the phase interval 0.31-0.37.
We then subtracted it from the from the average spectrum in the phase interval 0.38-0.60.
The difference spectrum shows modulations of the continuum also reminiscent of cyclotron 
harmonics but with much wider separation than those of the main pole (Fig. \ref{f:cyc2pole}). They seem to
be centred at $5.77\times 10^{14}$\,Hz and $7.81 \times 10^{14}$\,Hz,
respectively, and could be identified with the third and fourth cyclotron
harmonics. The identification is tentative, a model computation assuming
a field of 69 MG, which is also shown in the figure gives some support
for our interpretation  but needs confirmation. If confirmed this
measurement would give support to \SEPH's hypothesis of a non-dipolar
field structure. 

\begin{figure}
 \includegraphics{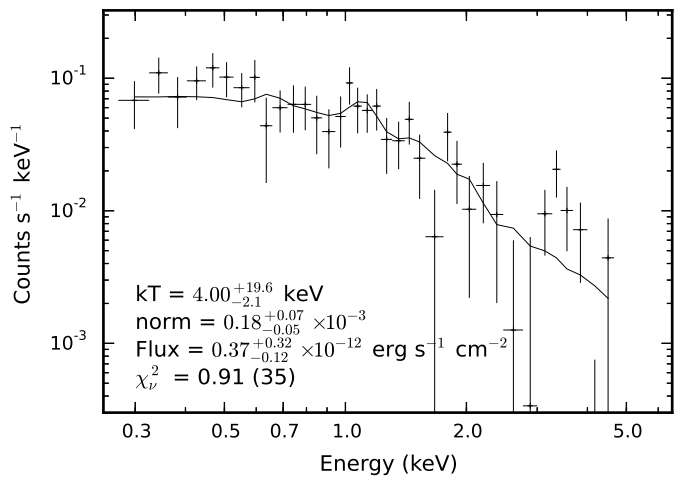}
 \caption{EPIC-$pn$ spectrum of the secondary hump ($0.45<\phi<0.65$) in Obs2, fitted with a single temperature {\sc Mekal}. The units of
 normalisation are the same as in Table \ref{tab:spectral_fits}. } 
 \label{fig:hump_spectrum}
\end{figure}

An \pn\ spectrum was extracted for the phase interval 0.45-0.65. The hump spectrum was 
successfully fitted with a single temperature {\sc Mekal} model with a plasma temperature of order 4 keV between 0.2 and 5.0~keV. 
The result is shown in Fig.~\ref{fig:hump_spectrum}.

The X-ray flux of the secondary hump is an order of magnitude smaller than 
that of the bright phase, and its optical flux is approximately $1.4\times 10^{-12}$
erg cm$^{-2}$ s$^{-1}$.
The optical and X-ray spectra, together with the shape and phasing 
of the secondary hump in the light curve, reveal a secondary accretion region 
centred at phase $\sim$0.58, i.e.~away from the mass-donating star.
As observed in several other polars with a two-pole accretion geometry, for instance RX~J1846.9+5538 \citep{SchwarzEtAl2002},
the weaker accreting pole has a significantly higher magnetic field strength. It is offset by only $\sim 140^\circ$
from the primary pole, which suggests a significantly off-centre WD magnetic field.

Since the secondary hump has duration $\Delta \phi_\text{hump}<0.5$, it must be located in the WD's
southern hemisphere. Assuming a point-like emission region, the approximation for the colatitude
\begin{equation}
 \beta = -\text{atan}\left[ \dfrac{1}{\cos\left(\pi\Delta\phi\right)\tan i}\right]
 \label{eqn:colatitude}
\end{equation}
is valid. From Fig. \ref{fig:OM_pf_lightcurves} we estimate the secondary hump to have $\Delta \phi_\text{hump}=0.2\pm 0.075$ and,
given $i=81.5^\circ \pm 2.2^\circ$ \SEPHp, we obtain colatitude $\beta_\text{hump}=169^\circ\pm5^\circ$. For high inclination $i$ and
small $\Delta \phi$, Eq. \ref{eqn:colatitude} is rather insensitive to changes in both, so our result is robust despite the
difficulties in observing the second bright spot. The secondary bright spot is therefore likely to be a few degrees nearer the WD rotation
axis than the primary bright spot ($\beta_\text{bright}\approx 18^\circ$).

\subsection{Eclipse parameters}
\label{sec:ephemeris}
We used the OM data and X-ray data to derive eclipse timings for
both observations. For the OM the times of ingress and egress
were taken to be times of half-intensity, found using a semi-automated version
of the ``cursor'' method described in \cite{SchwopeThinius2014} on 2~s binning. For the X-ray
data, we simply used the times of the corresponding Bayesian Block change points. The
results are summarised in Table \ref{tab:eclipse_timings}. We did not use
the OM data for the first and third eclipses in Obs1, because of data gaps.
For the last eclipse in Obs2 did use MOS2 data because of the buffer overflow
data gaps in the \pn\ data caused by the proton flare \citep{BurwitzEtAl2004}.

\begin{table*}
\caption{Timings of eclipse centres BJD(TDB) for \XMM\ optical monitor and
  EPIC-$pn$ instruments. To save space, the initial two digits (24) have
  been omitted from the timings. Offsets compared with the ephemeris of
  \SEPH\ are given in seconds. The cycle numbers are given according to the
  ephemeris in \SEPH. For the last cycle of Obs2, we have used the EPIC-MOS2 data
  because of data interruptions in EPIC-$pn$. Weighted averages for eclipse
  offsets and durations are given in the final row.} 
\label{tab:eclipse_timings}
 \begin{tabular}{llrllrl}
 Cycle  & OM & $\Delta t_\text{OM}$ (s) & Dur. (s) &EPIC-$pn$ & $\Delta t_\text{pn}$ (s)& Dur. (s)\\
 \hline
 14683 & ---                                &               &                &$56028.06349\pm2.1\times 10^{-5}$ & $-6.2\pm2.3$ & $434.6\pm3.6$\TopStrut \\
 14684 & $56028.14491\pm 1.3\times 10^{-5}$ & $-2.7\pm 1.8$ & $433.0\pm2.2$  &$56028.14488\pm2.2\times 10^{-5}$ & $-5.0\pm2.4$ & $433.8\pm3.7$\\
 14685 & ---                                &               &                &$56028.22632\pm1.9\times 10^{-5}$ & $ 0.1\pm2.2$ & $428.3\pm3.3$\\
 14686 & $56028.30766\pm 7.4\times 10^{-5}$ & $-3.0\pm 6.5$ & $441.6\pm12.7$ &$56028.30763\pm1.8\times 10^{-5}$ & $-5.0\pm2.1$ & $433.9\pm3.2$\\
 \hline
 19145 & $56391.16686\pm 1.0\times 10^{-5}$ & $-1.7\pm 1.8$ & $434.2\pm1.7$  &$56391.16686\pm1.0\times 10^{-5}$ & $-2.5\pm1.8$ & $435.7\pm1.7$\\
 19146 & $56391.24826\pm 1.9\times 10^{-5}$ & $0.1 \pm 2.3$ & $434.0\pm3.3$  &$56391.24820\pm1.0\times 10^{-5}$ & $-5.6\pm1.8$ & $440.8\pm1.7$\\
 19147 & $56391.32961\pm 1.7\times 10^{-5}$ & $-2.3\pm 2.1$ & $435.0\pm2.9$  &$56391.32961\pm1.5\times 10^{-5}$ & $-2.6\pm2.0$ & $435.8\pm2.5$\BottomStrut\\
 \hline
 average & ---                              & $-1.8\pm 0.5$ & $434.0\pm0.5$  & ---                              & $-3.4\pm0.9$ & $434.4\pm1.0$\BottomStrut\\
 \end{tabular}
\end{table*}

The eclipse centres in X-rays occur $3.4\pm1.0$ seconds earlier than predicted by 
the ephemeris in \SEPH, along wiht a smaller offset of $1.8\pm 0.5$~s in the OM data. 
The eclipses may be slightly longer by $0.7\pm 0.5$ and $1.1 \pm 1.0$~s in UV and X-rays, but
the differences are not statistically significant. The differences in the timings may indicate
that the X-ray and optical/UV emission do not arise from the same site in the accretion stream. 

\begin{figure}
\includegraphics{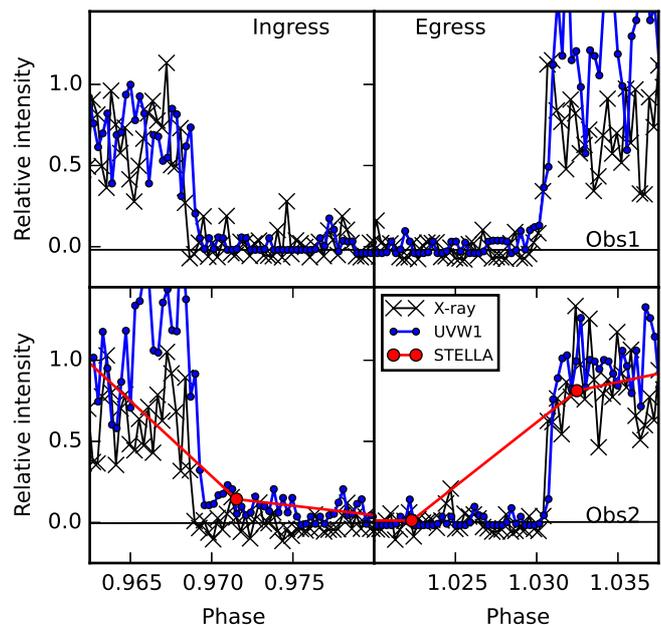}
\caption{Phase-folded eclipse profiles in \pn\ X-rays, UVW1, and STELLA optical wavelengths. The
binning is 2~s for X-rays and UVW1, and 60~s in STELLA.
The intensity is relative to the mean pre- and post-eclipse brightness. There is no evidence
of an extended emitting spot at any wavelength.}
\label{fig:eclipse_shapes}
\end{figure}

In Fig. \ref{fig:eclipse_shapes} we show the phase-folded eclipse profiles in X-ray, 
UV, and STELLA optical wavelengths for both observations. No STELLA observations were 
available for Obs1. The X-ray and UV observations are binned into 2~s intervals, and the
STELLA binning is 60~s. There is no indication that any of the eclipse ingresses or
egresses are resolved at these bin sizes, suggesting that the emitting spot is less than
2~s in extent at UV and X-ray energies.

\subsection{The SED and the energy balance}
\label{sec:energybalance}

According to the semi-empirical donor sequence of \cite{Knigge2006, Knigge2007}, the 1.953-hour orbital period implies 
a secondary of spectral type M4.6. Combining the $M_V=13.02$ and $M_B=14.44$ absolute magnitudes for this spectral type
with the observed $m_V\approx 21$ and $m_B\approx 22.4$ magnitudes during eclipse at low accretion rate, in \cite{ThorneEtAl2010} we obtained a distance to
the system of about $D=390$~pc. Assuming this distance and the
{\sc Mekal}+\emph{bbodyrad} bright phase spectral model, then the total bright
phase accretion luminosity $L_X=(\pi f_\text{bb} + 2\pi f_\text{mekal} + \pi f_\text{cyc})D^2$ was
$>3.9\times 10^{31}$ and $\approx 6.8\times 10^{31}$~erg~s$^{-1}$ for the two
observations. We conservatively set the accretion luminosity equal to the X-ray luminosity in Obs1 because
we have no optical data to allow the cyclotron component to be measured.
We infer accretion rates of $1.2\times 10^{-11}$ and $3\times
10^{-11} M_\odot$~yr$^{-1}$. These values are similar to the intermediate
accretion state of HU Aqr \citep{SchwarzEtAl2009}. 

We attempted to determine whether \CSS\ was detectable during eclipse by counting
source and weighted background photons between $0.975<\phi<1.025$ for all three
instruments. In both observations the photon counts were consistent with zero,
and we obtained 3$\sigma$ upper limits of 0.03 photons per second in both observations. Assuming
a {\sc Mekal} spectrum of temperature 1.0~keV, this puts an upper limit on the
X-ray luminosity during eclipse of $6.7\times 10^{31}$~erg~s$^{-1}$, or 4.55 times the
bolometric luminosity of the secondary. Because the X-ray luminosity of the secondary
is likely to be $\sim 1000$ times fainter, this result does not usefully constrain the luminosity of the parts of the
accretion stream unobscured during eclipse.

The cyclotron spectral component lying in the optical and UV regime is clearly visible in our observations, as is the bremsstrahlung component in X-rays. A third feature caused by
the absorption and re-radiation of some of the bremsstrahlung by the WD photosphere is expected as a blackbody-like bump at $\sim 20-50$~eV \citep{Cropper1990}, but
the \XMM\ observations do not extend to these low energies. The low energy blackbody component for Obs2 described in Table \ref{tab:spectral_fits} is underluminous if it
represents the re-emitted bremsstrahlung. We have attempted to place limits on an unobserved blackbody component for Obs2 lying in the hard UV/soft X-ray region in a
manner similar to \cite{RamsayCropper2002}.

\begin{figure}
 \includegraphics{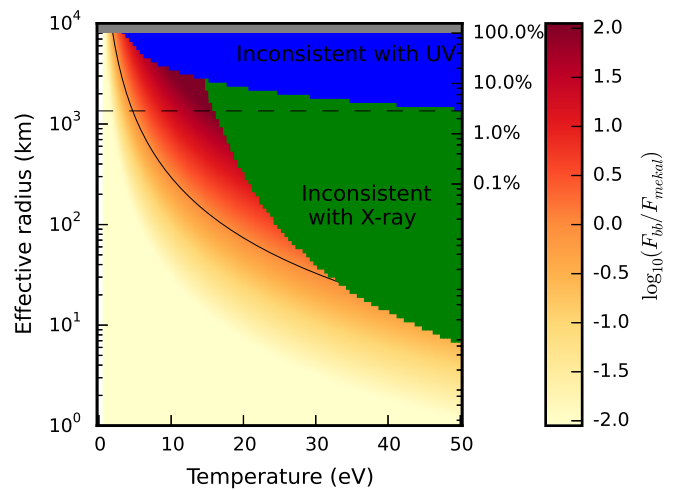}
 \caption{Parameter space in which a soft excess can exist. The blue and green regions indicate areas excluded because
 they are not consistent with the \XMM\ OM or X-ray observations, and the grey region indicates an emitting area larger than the 8000~km WD.
 Shaded points show the flux of the corresponding blackbody divided by the flux of the {\sc Mekal} component, and the black line
 tracks blackbodies with $F_{bb}=F_{mekal}$. The wedge-shaped region above the black line represents the area of the parameter space
 where a soft excess is still consistent with \XMM\ observations. The dashed horizontal line represents a circular spot with an eclipse ingress time of 10~s,
 the largest still consistent with the UVM2 ingress. The right $y$-axis expresses the emitting area as a percentage of the WD surface.}
 \label{fig:hiddenbb}
\end{figure}

The unobserved blackbody is subject to these restrictions: its emitting area can be no larger than the WD itself ($R_{bb}<8000$~km), it can be no cooler
than the WD ($T_{bb}>0.64$~eV for a very cool polar primary; \citep{SchmidtEtAl2005, FerrarioEtAl2015}), its energy density at UV wavelengths cannot exceed
the OM observations, and it must not cause the X-ray spectral fits to deteriorate. We generated a series of blackbodies with temperatures and normalisations
on a grid, with 0.75~eV~$\leq T_{bb} \leq$~50~eV and 
1~km~$\leq R_{bb}\leq8000$~km), where we have expressed the normalisation as the radius of a sphere 390~pc away. We excluded trials in which the energy density
exceeded the difference between bright and faint (excluding the secondary hump) phases in UV. We then fitted the Obs2 X-ray data between 200~eV and 1~keV with the 
{\sc Mekal}+\emph{bbodyrad} model of Table \ref{tab:spectral_fits}, replaced the original blackbody's parameters with those of the grid, and rejected trials that caused the $\chi^2_\nu$ to increase by more than 1.
Finally, we compared the fluxes of the remaining trials to the flux in the {\sc Mekal} component. The results are shown in Fig \ref{fig:hiddenbb}. There is a small
region of the parameter space in which a large soft excess could still exist. The cusp at $T_{bb}\approx 14.5$~eV, $R_{bb}\approx 2720$~km
accommodates a very large soft blackbody component ($F_{bb}\approx 380 F_{mekal}$), but this region is only barely consistent with the X-ray and UV observations, and it is
likely that better observations or more restrictive criteria would eliminate it. 

For example, if it is assumed that all of the bright-phase UV flux is the reprocessed blackbody then the UVM2 eclipse ingresses and egresses ($\leq10$~s; the low photon counts in the
UVM2 filter do not allow as accurate a determination as in UVW1) limit its emitting area to a region no larger than 6\% of the
observer-facing WD hemisphere, represented by the dashed black line in Fig. \ref{fig:hiddenbb}. This does not intersect the
blue region for cooler blackbodies, so in that area of parameter space, a significant portion of the UV flux must arise
from something other than the blackbody (most likely the cyclotron component) and that would in turn push the lower edge of the blue region
to smaller effective radii. The similarity of the UV and optical light curves suggests that most of the UV flux actually does come from cyclotron emission.
Though not excluded entirely, a significant soft excess can only be concealed in a very restricted area of the parameter space.

\section{Conclusions}
\label{sec:discussion}

We have performed a spectral and timing study of two \XMM\ observations of the
eclipsing polar \CSS. The star was brighter in X-rays and UV in the second
observation by a factor of approximately 2. \CSS\ has an X-ray hardness ratio of 
approximately 0.0-0.2 in the bright phase. These values are
similar to those found for HU Aqr by \cite{SchwarzEtAl2009} (though calculated
in slightly different energy bands), another point of similarity between these
two systems.

There is some evidence that the X-ray eclipses may lead the
optical and UV eclipses, suggesting that the X-ray plasma emission and cyclotron emission originate
in different sites along the accretion arc. This view is supported by our finding that $F_\text{cyc}\approx 3F_\text{mekal}$
for Obs2, hinting at a structured accretion region with parts being cyclotron-dominated \citep{Beuermann2004}.

The longitude of the bright phase changes from a trailing
spot during the lower accretion state to a leading one in the high accretion
state. A similar effect was observed by \SEPH\ but they caution that it
could have been due to observing the system with different filters at
different times. We can now rule out that explanation, because the \XMM\ observations show
the longitude shift occurring in X-rays and UV simultaneously. 

A bright phase dip in optical wavelengths at $\phi\approx 0.80-0.85$ has been
intermittently detected for this source and interpreted as partial obscuration of the
accretion spot by the stream of accreting material (\SEPH). It has not been observed
in the low accretion state but is usually detected in the intermediate and high
states. The exception to this rule is the 2013 April observation by the STELLA
telescope when, no dip was observed, despite being in the high state. Obs2 was
nearly concurrent with them, and we see no evidence in UV
of a bright phase dip, but obvious evidence for them in soft X-rays $\phi\approx 0.82$.
This behaviour is puzzling at first glance. However, as pointed out by \cite{WatsonEtAl1989, WatsonEtAl1990} in a study of EF Eri, the
dip in X-rays is due to photoelectric absorption, and in infrared/optical it is due to free-free absorption.
These two processes scale the optical depth as $N_H$ and $N_H^2$ respectively. Since we have found a dip column density
($N_H\lesssim 1.4\times 10^{21}$~cm$^{-2}$) at least an order of magnitude lower than for EF Eri ($\sim 4\times 10^{22}$~cm$^{-2}$) it follows that the bright phase 
dip ought to be two or more orders of magnitude less prominent for \CSS\ as for EF Eri.

Occurring just before eclipse ingress, the pre-eclipse dip is more perplexing. It implies that gas is
lost from the accretion stream very near the limb of the companion. However, we found that the bright phase as 
a whole is consistent with zero or very low absorption, suggesting that a negligible amount of absorbing gas is lost from the accretion stream 
between the two regions responsible for the two dips. Further study will be necessary to understand the behaviour of both absorption dips.

The ROSAT X-ray observatory viewed the sky position of \CSS\ for a total 
exposure of 192 seconds during its All-Sky Survey campaign, but \CSS\ was
not detected. If the system had been emitting as it was during the bright
phases of Obs1 or Obs2, ROSAT would have detected it with count rates of 0.17
and 0.49 counts per second between 0.1 and 2.4~keV. These count rates would 
have easily been sufficient to put it in the ROSAT bright source catalogue 
\citep{VogesEtAl1999}. Given that the 192 seconds of exposure were
accumulated over several scans that were separated by 96\,min, it is implausible
that the source could have escaped detection had it been in its
high accretion state. We conclude that it was in a low state during the RASS. 
It is interesting to note that all ROSAT-discovered polars showed a pronounced
soft blackbody-like radiation component, but bright polars without a prominent 
soft component have been discovered only in the \XMM\ era. The underlying
population of X-ray spectra will only be uncovered by the all-sky X-ray surveys
with eROSITA, which will be more sensitive than ROSAT in its own spectral band by
a factor of $\sim$30 \citep{MerloniEtAl2012, SchwopeEtAl2012}.

The WD in \CSS\ probably accretes onto both poles during the high accretion
state, because it shows a secondary bright phase in the UV and optical. The second bright
spot is faint but visible in X-rays, and its X-ray spectrum is consistent with a cooling accretion plasma.
Its optical spectra indicate a magnetic field twice as strong as the primary accretion spot, and it is not
located directly opposite the main accreting pole, suggesting a highly non-dipolar field geometry. Complex magnetic field configurations
seem to be common for polars (e.g. \citealt{BeuermannEtAl2007}). The secondary bright phase's position on the WD
hemisphere away from the mass-donor explains the low accretion rate but the trajectory used by matter to reach the remote pole
is not known. Determining how it occurs will provide a better understanding of these systems.

\begin{acknowledgements}
We gratefully acknowledge support for this project by the German DLR under
contract 50 OR 1405. This research made use of Astropy, a community-developed
core Python package for Astronomy (\citealt{Astropy2013}). We acknowledge with thanks the variable and comparison star
observations from the AAVSO International Database contributed by observers worldwide and used in this research. This study is
based partly on data obtained with the STELLA robotic telescope in Tenerife, an AIP
facility jointly operated by AIP and IAC. This work made use of data collected at the Calar Alto
observatory. We are grateful to the anonymous referee, whose comments improved the clarity of this work.
\end{acknowledgements}

\bibliography{cssbib.bib}

\end{document}